\documentstyle[aps,prl,eqsecnum,preprint,tighten]{revtex}

\begin{document}
\def\bq{\begin{equation}}
\def\eq{\end{equation}}
\title
{
AN EXTENSION OF LEVEL-SPACING UNIVERSALITY
 }
\author
{ E. Br\'ezin$^{1}$ and S. Hikami$^{2}$} 
\address
{$^{1}$ Laboratoire de Physique Th\'eorique, Ecole Normale
Sup\'erieure, 
24 rue Lhomond 75231, Paris Cedex 05, France{\footnote{
Unit\'e propre du centre national de la Recherche Scientifique, Associ\'ee
\`a l'Ecole Normale Sup\'erieure et \`a l'Universit\'e de Paris-Sud} 
}\\
$^{2}$ Department of Pure and Applied Sciences, University of Tokyo\\
{Meguro-ku, Komaba, Tokyo 153, Japan}\\
}
\maketitle
\begin{abstract}
In the theory of random matrices, several properties are known to be
 universal, i.e. independent of the specific probability distribution.
 For instance Dyson's
 short-distance universality of the correlation functions implies the
 universality of $P(s)$, the level-spacing distribution. We first briefly
 review how this property is understood for  unitary invariant ensembles
 and consider next a Hamiltonian $ H = H_0+ V $, in which $ H_0$ is a
 given, non-random, $N$ by $N$ matrix, and $V$ is an Hermitian random matrix
with a Gaussian probability distribution. The standard techniques, based on 
orthogonal polynomials, which are the key for the understanding of the
$H_0 = 0$ case,  are no longer available. Using then a completely
different approach, we 
derive closed expressions for the n-point correlation functions, which
are exact for finite N. Remarkably enough the result may still be expressed
as a determinant of an $n$ by $n$ matrix, whose elements are given by a kernel
 $K(\lambda,\mu)$ as in the $H_0=0$ case. From this
representation we can show that Dyson's short-distance universality 
still holds. We then conclude that $P(s)$ is independent of $H_0$.
\end{abstract}
\newpage
\section{INTRODUCTION}
 
Many years ago Wigner \cite{Wigner} introduced
 the level-spacing probability distribution 
$P(s)$, in his discussion of nuclear energy levels. 
The exact form of $P(s)$ was found later in the theory of random
matrices for the Gaussian unitary ensemble (GUE)   
\cite{Mehta,MehtaClo,Dyson}.
  This level-spacing probability distribution $P(s)$ 
  was empirically found to be universal in many different cases, for instance
 non-Gaussian probability distributions, or band matrices (in  which case
 the measure is not unitary invariant), and even for problems of an a priori
different nature such as  the 
level spacing of  the zeros of the Riemann zeta 
function \cite{Mehta,Mongo,Odlytzko}which is known to coincide with that of
 the GUE.
 
  In the next section, we first review how  the universality  of $P(s)$ 
has been derived
 for non-Gaussian unitary invariant ensembles, in which the probability
 measure  is given by
\bq\label{1.1}
 P(H) = {1\over{Z}}e^{ -{N} {\rm Tr} f(H) }\eq
where $f(x)$ is an arbitrary polynomial. One first integrates out
the unitary group, in order to obtain a probability distribution
 for the eigenvalues of $H$. It is then easy to show that the n-point
 function may be written as an $n$ by $n$ determinant; the matrix elements of
 this determinant are given by a kernel expressed in terms of orthogonal
 polynomials with respect to the weight ${\rm exp}[-N f(x)]$. Then the understanding
 of the relevant asymptotic behavior of these polynomials at large order
allows one to prove the short-distance universality of this kernel. From 
there one can derive the universality of $P(s)$ in the scaling limit in
which $N$ goes to infinity, the distance $x$ between two
 neighboring eigenvalues,
goes to zero, and  $s= N x$ is held fixed.

In the third section  we  consider a  Hamiltonian which is the sum of a given
 deterministic part $H_0$ and of a random  potential $V$ with a
 Gaussian probability distribution. The measure is not unitary invariant, but
one can still  write the probability distribution for  the eigenvalues
 of $H$ through  the well-known
Itzykson-Zuber integral \cite{Itzykson}. Generalizing a method  introduced
by  Kazakov \cite{Kazakov} for the density of eigenvalues, we write a 
representation of the n-level correlation function, in terms of an exact
 and explicit integral over $2n$ variables.  Then one discovers that an
 amazing algebraic structure allows one to express again  this n-point function
in terms of a determinant of an $n$ by $n$ matrix. The matrix elements
are given  by a kernel which has an explicit representation as an
integral over two variables. 
In a previous paper \cite{BH1}, we had discussed already the two-level
 correlation function
of this Hamiltonian through the same method,
 and we had  shown that the  behavior of this correlation function
is indeed universal, i.e. independent of the Hamiltonian $H_0$, in the
 short range scaling  limit, in which the distance $x$ of the  two energy
 levels
 becomes small, and $N$ goes to infinity, with fixed $Nx$. We had also 
briefly discussed the n-point function in \cite{BH2}. The main steps are
recalled here;  the universality of $P(s)$ follows immediately.

In the last section we establish some properties of this kernel, and show that
it does satisfy some necessary consistency conditions.

\section{ LEVEL-SPACING DISTRIBUTION $P(s)$ FOR 
GENERALIZED GUE ENSEMBLES
 }

 We return to the single random matrix case with a probability 
\bq\label{2.1}
 P(H) = {1\over{Z}}e^{ -{N} {\rm Tr} f(H) }\eq
and integrate out the unitary degrees of freedom.
 The resulting probability distribution for the $N$ eigenvalues
 of $H$ is \cite {Mehta} 
\bq\label{2.2}
  P_N(x_1,\cdots,x_N) = C \prod_{i<j} ( x_i - x_j )^2 e^{ -
 {N} \sum_{i=1}^{N}f(x_i)} 
\eq
The n-point 
correlation function $R_n(x_1,...,x_n)$, is defined as 
\bq\label{2.3}
   R_n(x_1,...,x_n) ={N!\over{(N-n)!}}\int_{-\infty}^{\infty} \cdots 
\int_{-\infty}^{\infty} dx_{n+1}\cdots dx_N 
P_N(x_1,\cdots,x_N)
\eq
Following Mehta \cite{Mehta}, we introduce the  orthogonal polynomials
$\phi_k(x)$ with 
respect to the measure ${\rm exp}[-Nf(x)]$. Then  $R_n$ is given 
by the determinant,
\bq\label{2.4}
    R_n(x_1,\cdots,x_n) = {\rm det} [K_N(x_i,x_j)]_{i,j=1,...,n}.
\eq
in which the kernel $K_N(x,y)$ is  expressed as a sum of orthogonal
polynomials
\bq\label{2.5}
    K_N(x,y) = {1\over{N}} e^{ -
 {N\over{2}}(f(x)+ f(y)) }\sum_{k=0}^{N-1}
\phi_k(x)\phi_k(y).\eq
For instance the pair correlation function, the $n = 2$ case,  
becomes
\bq\label{2.6}
   R_2(x_1,x_2) = \rho(x_1)\rho(x_2) - K_N(x_1,x_2)K_N(x_2,x_1)
\eq
in which the density of states  $\rho(x)$ is the diagonal part of the kernel
 $\rho(x) = K_N(x,x)$. With our normalization conventions the
 density of state $\rho(x)$
has a support of finite extension in the large N limit.

In the short distance scaling limit, $K_N(x_1,x_2)$ becomes
 \cite{Mehta,Dyson,BZ1}
\bq\label{2.7}
    K_N(x_1,x_2) \simeq {\sin[ \pi N (x_1 - x_2) \rho({(x_1 + x_2)
\over{2}})]\over{\pi 
N (x_1 - x_2 )}}
\eq
for $N\rightarrow \infty$, $x_1 - x_2 \rightarrow 0$ and finite $N(x_1-x_2)$.
 The universality of (\ref{2.7}) with respect to the function $ f(x)$ which 
characterizes the probability measure is thus manifest. The universality of the level-spacing distribution $P(s)$ follows at once.

Indeed, following Mehta \cite {Mehta}, we first compute the probability
 $E(\theta)$ 
that the interval $[-{\theta\over{2}},{\theta\over{2}}]$ does not 
contain any of the points $x_1,...,x_N$ in the large N limit. It is thus
 obtained
by  integrating the N variables of  $P_N(x_1,...,x_N)$ outside  the interval
$[-{\theta\over{2}},{\theta\over{2}}]$ :
\bq\label{2.8}
  E(\theta) = \int_{out} \cdots \int_{out} P_N(x_1,...,x_N)dx_1 \cdots dx_N
\eq
where the integrals are performed outside the region 
$[-{\theta\over{2}},{\theta\over{2}}]$;
\bq\label{2.9}
   \int_{out}dx = (\int_{-\infty}^{\infty} - 
\int_{-{\theta\over{2}}}^{{\theta\over{2}}})dx
\eq
We may thus express $E(\theta)$ in terms of the  $R_n$'s by using
 systematically (\ref {2.9}) for all the N variables:
\bq\label{2.10}
    E(\theta) = 1 - N \int_{-{\theta\over{2}}}^{{\theta\over{2}}} \rho(x) dx 
+ {{N}^2\over{{2}!}} \int_{-{\theta\over{2}}}^{{\theta\over{2}}}
\int_{-{\theta\over{2}}}^{\theta\over{2}}
R_2(x,y)dx dy + \cdots.
\eq
The natural scale for the  level spacing $\theta$ is of  order
 ${1\over{N}}$ since in the large N limit the support  
 of the density of state is finite. 
We thus consider the short distance scaling limit, in which 
 $\theta$ goes to zero and N to  infinity, with fixed   $N\theta$.
In that scaling limit
\bq\label{2.11}
   N \int_{-{\theta\over{2}}}^{{\theta\over{2}}} \rho(x) dx 
= N\theta\rho(0) + O(1/N).
\eq 
We thus define the scaling variable
\bq\label{2.12}
 N\theta \rho(0) = s.
\eq 
 The next terms of (\ref{2.10}) are  obtained  
in this limit by the change of variables
\bq\label{2.13}
 N x \rho(0) = x^{'}.
\eq 
Then, in the scaling limit,
\bq\label{2.14} 
 {{N}^2} \int_{-{\theta\over{2}}}^{{\theta\over{2}}}
\int_{-{\theta\over{2}}}^{\theta\over{2}}
R_2(x,y)dx dy  = \int_{-{s\over{2}}}^{{s\over{2}}}
\int_{-{s\over{2}}}^{s\over{2}}
{\tilde R}_2(x',y')dx' dy' 
\eq 
with
\bq\label{2.15}
    {\tilde R}_n(x_1,\cdots,x_n) = {\rm det} [{\tilde K}(x_i,x_j)]_{i,j=1,...,n}.
\eq
in which 
\bq\label{2.16}
   {\tilde K}(y_1,y_2) = {{\sin[ \pi (y_1 - y_2) )]}
\over{\pi  (y_1 - y_2 )}}.
\eq  
In this scaling limit we thus obtain 
\bq\label{2.17}
   E(s) ={\sum_{n=0}^{\infty}} {{(-1)}^n\over{{n}!}} 
\int_{-{s\over{2}}}^{{s\over{2}}}
\cdots\int_{-{s\over{2}}}^{{s\over{2}}} dx_1,\cdots,dx_n 
 {\rm det} [{\tilde K}(x_i,x_j)]_{i,j=1,...,n}
\eq

  From this representation it is easy to expand $E(s)$ for $s$ small;
 for instance
\bq\label{2.18} 
  \int_{-{s\over{2}}}^{{s\over{2}}}
\int_{-{s\over{2}}}^{s\over{2}}
{\tilde R}_2(x,y)dx dy =  {\pi^2\over{36}} { s}^4 - {\pi^4\over{675}} { s}^6
 + O( {s}^8) 
\eq
and since the $n=3$ term of (\ref{2.17}) is easily shown to be 
 of order ${s}^7$ for s small, we
find
 \bq\label{2.19} 
 E(s) = 1 - s  + {\pi^2\over{36}} { s}^4 - {\pi^4\over{675}} { s}^6
 + O( {s}^7). 
\eq
One can also introduce the eigenvalues $\lambda_i(s)$ of the 
integral equation for the kernel $\tilde K$
 on the interval $[-{s\over{2}}, +{s\over{2}}]$,
\bq
    \int_{-{s\over{2}}}^{s\over{2}} \tilde K (x,y) \psi_i (y) dy = 
\lambda_i \psi_i(x).
\eq
From (\ref{2.17})
 we can write
\bq\label{2.20}
E(s) = \prod_{i=1}^{\infty} (1-\lambda_i) = {\rm det} [1 - {\tilde K}].
\eq
For  small $s$, a perturbational expansion using Legendre polynomial
gives the same result as (\ref{2.19}). 
The level-spacing probability distribution $P(s)$ is now  obtained from $E(s)$
\bq\label{2.21}
  P(s) = {d^2\over{ds^2}} E(s)
\eq

Through this representation, we find 
that the universality of $P(s)$ results  from 
two sources; i) the n-point correlation $R_n$ is expressed as the determinant
of a  kernel $K_N(x_i,x_j)$, ii) the kernel $K_N(x,y)$ has a universal short 
distance behavior  $\tilde K$ in the short distance scaling limit.

\section{ DETERMINISTIC PLUS RANDOM HAMILTONIAN }
We now consider a Hamiltonian $H = H_0 + V$, where $H_0$ is a given,
 non-random, $N \times N$ hermitian matrix, and 
$V$ is a random Gaussian hermitian matrix. 
The probability  distribution  $P(H)$ is thus given
by
\begin{eqnarray}\label{3.1}
P(H) &=& {1\over{Z}}e^{ - {N\over{2}} {\rm Tr} V^2 }\nonumber\\
&=&
{1\over{Z^{'}}}e^{- {N\over{2}}{\rm Tr} ( H^2 - 2 H_0 H)} 
\end{eqnarray}

We are thus  dealing with a Gaussian unitary 
ensemble modified by the external matrix source 
 $H_0$, which breaks the unitary invariance of the measure.
In  previous  work \cite{BH1,BH2}, we have already discussed 
 the density of state, 
and the two-level correlation function. For completeness, we repeat here
the basic steps.
The density of state $\rho(\lambda)$ 
 is
\begin{eqnarray}\label{3.2}
\rho({\lambda}) &=& {1\over{N}} < {\rm Tr} \delta ( \lambda - H ) >\nonumber\\
&=&
\int_{-\infty}^{+\infty} {dt\over{2 \pi}} e^{- iNt \lambda} U(t)
\end{eqnarray}
where $U(t)$ is the average "evolution" operator
\begin{equation}\label{3.3}
U(t) = < {\rm Tr} e^{iNt H} >. 
\end{equation}
We first integrate over the unitary matrix $\omega$ which 
diagonalizes $H$ in (\ref{3.1}),
and without loss of generality we may assume that $H_0$ 
is a diagonal matrix
with eigenvalues $(\epsilon_1, \cdots, \epsilon_N)$.
 This is done with the help of the well-known
Itzykson-Zuber integral \cite{Itzykson}, 
\begin{equation}\label{3.4}
\int
d{\omega} {\rm exp}( {\rm Tr} A {\omega} B {\omega} ^{\dag} ) =
{{\rm
det}({\rm exp}(a_i b_j))\over{\Delta(A)\Delta(B)}}
\end{equation}
where
$\Delta(A)$ is the Van der Monde determinant constructed with the
eigenvalues of A:
\begin{equation}\label{3.5}
\Delta(A) = \prod_{i<j}^N (a_i - a_j).
\end{equation}
We are then led to
\begin{eqnarray}\label{3.6}
U(t) = && {1\over{Z^{'}\Delta(H_0)}}\sum_{\alpha =
1}^{N} \int
dx_1 \cdots dx_N e^{i Nt x_\alpha} \Delta(x_1,\cdots,x_N) \nonumber\\
&&\times {\rm exp}( - {N\over{2}}\sum x_i^2 + N \sum \epsilon_i x_i ).
\end{eqnarray}
The normalization is fixed by
\bq\label{3.7}
U(0) = N
\eq
The integration over the $x_i$'s may be done easily, if we note that
\begin{eqnarray}\label{3.8}
&&\int dx_1 \cdots dx_N \Delta(x_1,\cdots,x_N) {\rm exp}( - {N\over{2}}\sum
x_i^2 + N \sum b_i x_i ) \nonumber\\
&&=
\Delta(b_1,\cdots,b_N) {\rm exp}({{N\over{2}}\sum b_i^2}) 
\end{eqnarray}

Putting 
$
b_i = \epsilon_i + i t \delta _{\alpha,i} 
$, we obtain
\begin{equation}\label{3.9}
U(t) =\sum_{\alpha=1}^{N} \prod_{\gamma\neq \alpha}^N
({\epsilon_\alpha - \epsilon_\gamma + i t\over{\epsilon_\alpha
 - \epsilon_\gamma}}) e^{-
{Nt^2\over{2}} +N i t \epsilon_\alpha }
\end{equation}
The sum over N terms in (\ref{3.9}) 
may then be replaced by a contour-integral in the complex 
plane,
\begin{equation}\label{3.10}
U(t) = {1\over{i t}} \oint {du\over{2 \pi i}} \prod_{\gamma = 1}^{N}
({u - \epsilon_\gamma + i t\over{ u - \epsilon_\gamma}})
 e^{- {N t^2\over{2}} + i t N u  }
\end{equation}
The contour of integration encloses all the eigenvalues $\epsilon_\gamma$. 
The Fourier transform with respect to $t$ gives  the density of state
 in the presence of an 
arbitrary external source $H_0$ and for finite $N$.

In the case of  the two-point correlation function, we have  
\bq\label{3.12}
R_{2}(\lambda,\mu) = <{1\over{N}} {\rm Tr}(\lambda - H)
{1\over{N}}{\rm Tr}
(\mu - H ) >
\eq
By using integral representations for the two $\delta$-functions,
the two-point correlation function $R_2(\lambda,\mu)$ is 
expressed  as the Fourier transform of $U(t_1,t_2)$,
\begin{equation}\label{3.13}
U(t_1,t_2) = <{\rm Tr} e^{iN t_1 H}  {\rm Tr} e^{iN t_2 H} >
\end{equation}
Using again the Itzykson-Zuber formula to integrate over  the unitary group,
 we obtain
\bq\label{3.14}
U(t_1,t_2) =
\sum_{\alpha_1,\alpha_2 = 1}^N \int \prod_{i = 1}^N
dx_i {\Delta(x)\over{\Delta(H_0)}}e^{-N\sum ({1\over{2}}x_i^2 -
 x_i \epsilon_i) + iN( t_1 x_{\alpha_1} + t_2 x_{\alpha_2})} \eq

After integration over the $x_i$'s, we have
\begin{eqnarray}\label{3.15}
&& U(t_1,t_2) = \sum_{\alpha_1,\alpha_2} {\prod_{i<j}(\epsilon_i -
\epsilon_j + i t_1(\delta_{i,\alpha_1} - \delta_{j,\alpha_1}) +
it_2(\delta_{i,\alpha_2}-
\delta_{j,\alpha_2}))\over{\prod_{i<j}(\epsilon_i - \epsilon_j)}}
\nonumber\\
&&\times
e^{ N i t_1 \epsilon_{\alpha_1} + 
N i t_2 \epsilon_{\alpha_2} -{N\over{2}} t_1^2 -
{N\over{2}}t_2^2 - N t_1 t_2 \delta_{\alpha_1,\alpha_2}}
\end{eqnarray}
This term is devided into two parts; $\alpha_1 = \alpha_2$  and 
$\alpha_1 \neq \alpha_2$ cases,
\begin{eqnarray}\label{double}
 &&U(t_1,t_2) = \sum_{\alpha_1} \prod_{i<j} 
{(\epsilon_i - \epsilon_2 + i (t_1 + t_2) 
(\delta_{i,\alpha_1}-\delta_{j,\alpha_1})
\over{(\epsilon_i - \epsilon_j)}}
 e^{N i (t_1+t_2) \epsilon_{\alpha_1}- {N\over{2}}(t_1 + t_2)^2}
\nonumber\\
&& + \sum_{\alpha_1 \neq \alpha_2}
{(\epsilon_{\alpha_1} - \epsilon_{\alpha_2} + i (t_1 - t_2))\over{\epsilon_{\alpha_1} - \epsilon_{\alpha_2}}}
\prod_{\gamma\neq(\alpha_1,\alpha_2)} {(\epsilon_{\alpha_1} - 
\epsilon_{\gamma} + i t_1)\over{\epsilon_{\alpha_1} - \epsilon_{\gamma}}}
{(\epsilon_{\alpha_2} - \epsilon_\gamma + i t_2)\over{\epsilon_{\alpha_2}
-\epsilon_\gamma}}
\nonumber\\
&&\times e^{N i t_1 \epsilon_{\alpha_1} + N i t_2 \epsilon_{\alpha_2} - 
{N\over{2}} (t_1^2 + t_2^2)}
\end{eqnarray}
Fourier transform of the first term becomes $\delta$-function \cite{BHZ2}, and
can be neglected for $R_2(\lambda,\mu)$ for $\lambda \neq \mu$.
The double sum in (\ref{double}) may be written again as an integral over two complex
 variables:
\begin{eqnarray}\label{3.16}
U(t_1,t_2) &=&  {1\over{(t_1t_2)}}e^{ -{N\over{2}} t_1^2 -
{N\over{2}}t_2^2} \oint {dudv\over{(2\pi i)^2}}
e^{N i t_1 u +N i t_2 v}
 {(u - v + (i t_1 - i t_2))(u - v) \over{(u - v +  i t_1)( u - v - i t_2)}
}\nonumber\\
&\times&\prod_{\gamma=1}^N 
( 1 + {it_1\over{(u - \epsilon_\gamma)}})( 1 + { it_2\over{
(v - \epsilon_\gamma)}}) 
\end{eqnarray}
Noting that 
\bq
    1 - {t_1 t_2 \over{( u - v + i t_1)( u - v - i t_2)}} 
= {( u - v + i (t_1 - t_2))( u - v)\over{( u - v + i t_1)( u - v - i t_2)}}
\eq
we find that (\ref{3.16}) is a sum of the disconnected term and a connected
part.
We know Fourier transform U with repect to $t_1$ and $t_2$ and shift the
 integrations variables.  By the shifts $t_1 \rightarrow t_1 + i u $, 
and $t_2 \rightarrow 
t_2 + i v $,  we see easily that $ R_2(\lambda, \mu)$ is a two by two
 determinatnt, namely that
\bq\label{3.30}
R_2(\lambda, \mu) = K_N(\lambda,\lambda) K_N(\mu,\mu) - 
 K_N(\lambda,\mu) K_N(\mu,\lambda)
\eq
with the kernel
\bq\label{3.17}
K_N(\lambda, \mu) =  \int {dt\over{2 \pi}}\oint 
{dv\over{2 \pi i}}\prod_{\gamma = 1}^{N}
( { \epsilon_\gamma - i t\over{ v - \epsilon_\gamma}})
{1\over{v - i t}} e^{- {N\over{2}}v^2 - {N\over{2}} t^2 
- N i t \lambda + N v \mu}
\eq
Note the similarity of the determinantal structure found here with
 that of the zero source case given in (\ref{2.4}).

 In \cite{BH1}, this kernel $K_N(\lambda,\mu)$ was examined in the 
 scaling limit, large N, but fixed $N(\lambda -\mu)$. In this limit one
 can evaluate the kernel (\ref{3.17}) by the saddle-point method.
 The result was found to be, up to a phase factor that we omit here,  
\bq\label{3.18}
K_N(\lambda_1,\lambda_2) = - {1\over{\pi y}} {\rm sin} [\pi y \rho(\lambda_1)]
\eq
where 
$y = N(\lambda_1 - \lambda_2)$.
Apart from the scale dependence provided by  the density of state $\rho$,
the two-point correlation 
function has  a  universal scaling limit, i.e. indepent of the deterministic
part $H_0$ of the random Hamiltonian.

\section{ DETERMINANT FOR THE n-POINT CORRELATION 
FUNCTION }

The n-point correlation function $R_n(\lambda_1,\cdots,\lambda_n)$ is 
given by
\bq\label{4.1}
   R_n(\lambda_1,\cdots,\lambda_n) = {1\over{N^n}}
<\prod_{i=1}^{n} {\rm Tr} 
\delta (\lambda_i - M)>
\eq
If we  put the constraints that all $\lambda_i$ are different, this 
expression conincides with (\ref{2.1}). When some $\lambda_i$ become
same, we have extra $\delta$-functions as shown in \cite{BH1}. 
Therefore, we assume all $\lambda_i$ are different.
  
Without an external source, this n-point correlation function 
is expressed  in terms of the kernel $K_N(\lambda_i,\lambda_j)$ as
\cite{Mehta,Dyson}
\bq\label{4.2}
   R_n(\lambda_1,\cdots,\lambda_n) = {\rm det} [K_N(\lambda_i, \lambda_j)]
\eq
where $i,j = 1, \cdots, n$.
This result  was derived  by the use of the orthogonal polynomials.
In the external source problem, we can not apply the orthogonal polynomial 
method. Our aim is to find a proof of (\ref{4.2}) for the external source case.

Using the Itzykson-Zuber formula of (\ref{3.5}), we have
\bq\label{4.3}
   R_n(\lambda_1,\cdots,\lambda_n) = {1\over{N^n}}\sum_{\alpha_i \neq \alpha_j} 
\int {dt_1 \cdots dt_n \over{(2 \pi)^n}}
{\Delta(B)\over{\Delta(H_0)}}
e^{{N\over{2}} \sum b_i^2 + i \sum t_k \lambda_k} 
\eq
where
\bq\label{4.4}
   b_k = \epsilon_k + i ( t_1 \delta_{k,\alpha_1} + \cdots + t_n 
\delta_{k,\alpha_n} ).
\eq

Using the contour-integration representation, we get
\begin{eqnarray}\label{4.5}
  R_n &=& \int {dt_1 \cdots dt_n\over{(2 \pi )^n}} 
e^{ - {N\over{2}} \sum t_p^2 + i N \sum t_p \lambda_p}\nonumber\\
&\times&
\oint {du_1 \cdots du_n\over{(2\pi i)^n}} e^{Ni\sum t_p u_p}
\prod_{p=1}^{n} \prod_{\alpha=1}^{N} ( 1 + {i t_p\over{ u_p - \epsilon_\alpha} })\nonumber\\
&\times& \prod_p^n {1\over{t_p}}
\prod_{p<q} {[ u_p - u_q + {i}(t_p - t_q)](u_p - u_q)
\over{(u_p - u_q + {i} t_p) ( u_p - u_q - {i} t_q)}} 
\end{eqnarray}
When $n = 2$, this reduces to the previous expression (\ref{3.16}).
We make a shift of the variables $t_p$:  $t_p\rightarrow 
t_p +  i u_p$
then we get
\begin{eqnarray}\label{4.6}
R_n &=& \int {dt_1 \cdots dt_n\over{(2 \pi )^n}} \oint
{du_1 \cdots du_n\over{(2 \pi i)^n}}
 e^{-{N\over{2}} \sum t_p^2 - {N\over{2}} \sum u_p^2 + \sum \lambda_p ( - 
iN t_p + N u_p)}\nonumber\\
&\times& \prod_{p=1}^{n} \prod_{\alpha=1}^{N} 
({ - \epsilon_{\alpha} + i t_p\over{   u_p - \epsilon_\alpha}})
\prod_{p<q} ({it_p - it_q
\over{- u_q + it_p}}) {( u_p - u_q)
\over{( u_p - it_q)}}\prod_{p=1}^n 
{1\over{(t_p + i u_p)}}
\end{eqnarray}

We recognize in (\ref{4.6})a  Cauchy determinant,
\bq\label{4.7}
 {\rm det} \left [{1\over{a_i - b_j}}\right ]_{i,j=1,...,n} = 
( - 1)^{{n(n-1)\over{2}}} 
{\prod_{i<j} (a_i - a_j) 
( b_i - b_j)\over{ \prod_{i,j}( a_i - b_j)}}
\eq
if we identify $a_k$ to  ${it_k}$, and $b_k$ to  $u_k$ in (\ref{4.6}).
Then, $R_n$ is given by 
\begin{eqnarray}\label{4.8}
R_n &=& \int {dt_1 \cdots dt_n\over{(2 \pi )^n}} \oint
{du_1 \cdots du_n\over{(2 \pi i)^n}}
 e^{-{N\over{2}} \sum t_k^2 - {N\over{2}} 
\sum u_k^2 + \sum \lambda_k (- i N t_k + N u_k)}\nonumber\\
&\times& \prod_{k=1}^{n} \prod_{\alpha=1}^{n} 
({ - it_k+ \epsilon_\alpha\over{\epsilon_\alpha - u_k}})
{\rm det} ( {1\over{ it_i - u_j}} )\nonumber\\
&=& \int {dt_1 \cdots dt_n\over{(2 \pi )^n}} \oint
{du_1 \cdots du_n\over{(2 \pi i)^n}}
 e^{-{N\over{2}} \sum t_k^2 - {N\over{2}} 
\sum u_k^2 + \sum \lambda_k (- i N t_k + N u_k)}\nonumber\\
&\times& {\rm det} \left [ \prod_{\alpha=1}^{N} { - it_i + \epsilon_{\alpha}
\over{(it_i - u_j)(\epsilon_{\alpha} - u_j)}}\right ]
\end{eqnarray}
 Using the expression for the kernel of (\ref{3.17}), we 
obtain
\bq\label{4.9}
R_n(\lambda_1,\cdots,\lambda_n) = {\rm det}\left [
K_N(\lambda_i,\lambda_j)\right ]_{i,j=1,...,n}
\eq
 We could thus prove the determinantal form of the n-point 
correlation function for a  deterministic plus random 
Hamiltonian.

\section{ THE PROPERTIES OF THE KERNEL }

 As we have seen in (\ref{4.9}), the n-point correlation function 
 $R_n$ is 
expressed by the determinant in the presence of the external source.
When we integrate out the variables $x_{l + 1},...,x_n$ of $R_n(x_1,
...,x_n)$, we obtain the l-point correlation function $R_l(x_1,...,x_l)$.
Since we have (\ref{4.9}), the necessary consistency condition for this result
is 
\bq\label{5.1}
   \int_{-\infty}^{+\infty} d\mu K_N(\lambda,\mu) K_N(\mu,\nu)
   = K_N(\lambda,\nu)
\eq
This property is verified easily by the contour-integral representation
of the kernel $K_N(\lambda,\mu)$
given in (\ref{3.17}) \cite{BH2}. We have
\begin{eqnarray}\label{5.2}
&&\int_{-\infty}^{\infty} K_N(\lambda,\mu)K_N(\mu,\nu) d\mu\nonumber\\
&&=\int_{-\infty}^{\infty}{dt_1 dt_2\over{(2 \pi)^2}}\oint
{du_1 du_2\over{(2 \pi i)^2}}\prod_{\gamma}
({\epsilon_\gamma + i t_1\over
{u_1 - \epsilon_\gamma}})({\epsilon_\gamma + i t_2\over{u_2 - 
\epsilon_\gamma}}){1\over{(u_1 + i t_1)(u_2 + i t_2)}}\nonumber\\
&&\times e^{-{N\over{2}} (u_1^2 + u_2^2 + t_1^2 + t_2^2) - iN t_1 \lambda
- i N t_2 \mu - N u_1 \mu - N u_2 \nu}.
\end{eqnarray} Integration over $\mu$, after the shift $t_2 \rightarrow 
t_2 + i u_1$,  gives  a delta-function for $t_2$, and the contour-integral
over  $u_1$ around the pole $u_1 = - i t_1$ reconstructs  $K_N(\lambda,\nu)$.

We observe also the kernel $K_N(\lambda,\mu)$ has  $N$ eigenvalues equal
 to one,
 with Hermite polynomials as eigenfunctions since, for  $n < N$,
\bq\label{5.3}
\int_{-\infty}^{\infty} K_N(\lambda,\mu) H_n(\sqrt{N} \mu)e^{-{N\over{2}}
\mu^2} d\mu = H_n(\sqrt{N}\lambda) e^{-{N\over{2}} \lambda^2}
\eq
with, $H_0(x)=1$, $H_1(x) $= $x$, $H_2(x)
$ = $x^2 - 1$, etc.
This  property may  also be easily verified through the contour 
integratal representation.
For $n > N - 1$, (\ref{5.3}) does not hold. The right hand side of 
(\ref{5.3}) becomes $\epsilon_\gamma$ dependent. When the external 
source $\epsilon_\gamma$ goes to zero, the right hand side of (\ref{5.3})
is vanishing for $n > N - 1$. This is of course related to the fact that
the kernel is then expressed as a finite sum of Hermite polynomials.

\section{SUMMARY AND DISCUSSION }

In the previous section, we have proved that the n-point correlation function
is expressed by the kernel $K(x,y)$ as in the absence of 
an external source. In the short distance limit, in which 
$(\lambda - \mu)N$ is kept fixed, the kernel $K_N(\lambda_p,\lambda_q)$ 
takes a  universal form, and the n-point correlation becomes 
universal ( up to a  rescaling by the density of state $\rho$).
As we have seen in section 2, the level-spacing probability
distribution $P(s)$ is given by an integration over the n-point 
correlation function $R_n(\lambda_i,...,\lambda_n)$. Therefore,
in the short distance scaling limit, 
$P(s)$ has a  universal form, independent of the deterministic 
part.

 We have assumed throughout this work  that the density of state is
 finite,of  order one,
in  the energy range that we are considering.
We have discussed the level-spacing probability $P(s)$ for two levels centered
 around the enregy zero. If we considered instead two levels
centered around an energy  $ E_0$, i.e. an interval $[ - {s\over{2}} + E_0,
 {s\over{2}} + E_0 ]$, the behavior of the kernel $K_N(\lambda,
\mu)$ remains  universal, apart from the scaling by the the density of state
$\rho(E_0)$instead of $\rho(0)$. Therefore, we still have the same 
universal spacing dsitribution  $P(s)$
for an  arbitrary energy $E_0$ as long as $\rho(E_0)$ remains of order 
one.
 
 We have also  assumed that the eigenvalues of the deterministic term
$H_0$ are inside the support of the asymptotic smooth density of state $\rho$.
When the eigenvalues are widely separated, the density of state
shows an oscillatory behavior. In such cases, the two-level
correlation function, or the kernel $K_N(\lambda,\mu)$ 
does not approach, in the scaling limit, the sine-kernel, 
and a  universal 
form for  $P(s)$ is not expected \cite{Guhr}. This is reasonable, since we 
know  that, when  the random potential $V$ increases in  comparison 
with the unperturbed  deterministic term $H_0$, we crossover 
to a  universal behavior 
independent of the initial deterministic term.
 
 Finally we have found two kinds of universality: either $H_0=0$ and
 the distribution of $H$ is  non-Gaussian, or $H_0$ is non-zero
 and $V$ is Gaussian. It is tempting to conjecture that this generalizes to
non-Gaussian problems with a non-zero source as  considered  in the case
 of the two-level correlation function 
\cite{BZ1,BZ2,BHZ1}, or to the time-dependent case \cite{BH2}.

\acknowledgements
This work was supported by 
the CNRS-JSPS cooperative project, and by the CREST of JST. 

\end{document}